\documentclass[preprint,review,1p]{elsarticle}
 \usepackage{amssymb}
 \usepackage{amsmath}
 \usepackage{amsthm}
\newtheorem{theorem}{Theorem}[section]

\newtheorem{remark}[theorem]{Remark}

\begin{document}

 \begin{frontmatter}

 \title{Derivation of the Camassa-Holm equations for elastic waves}

\author{H. A. Erbay$^{1}$\corref{cor1}}
    \ead{husnuata.erbay@ozyegin.edu.tr}
\author{S. Erbay$^1$}
    \ead{saadet.erbay@ozyegin.edu.tr}
\author{A. Erkip$^2$}
    \ead{albert@sabanciuniv.edu}
\cortext[cor1]{Corresponding author. Tel: +90 216 564 9489 Fax: +90 216 564 9057}

\address{$^1$ Department of Natural and Mathematical Sciences, Faculty of Engineering, Ozyegin University,  Cekmekoy 34794, Istanbul, Turkey}

\address{$^2$ Faculty of Engineering and Natural Sciences, Sabanci University,  Tuzla 34956,  Istanbul,    Turkey}

 \begin{abstract}
 \noindent
 In this paper we provide a formal derivation of both the Camassa-Holm equation and the fractional Camassa-Holm equation for the propagation of small-but-finite amplitude long waves  in a nonlocally and nonlinearly elastic medium. We first show that the equation of motion for the nonlocally and nonlinearly elastic medium reduces to the  improved Boussinesq equation for a particular choice of the kernel function appearing in the integral-type constitutive relation.  We then derive the Camassa-Holm equation from the improved Boussinesq equation using an asymptotic expansion valid  as nonlinearity and dispersion parameters tend to zero independently.  Our approach follows mainly  the standard techniques used widely in the literature to derive the Camassa-Holm equation for shallow water waves.  The case where the Fourier transform of the kernel function has fractional powers is also considered and the fractional Camassa-Holm equation  is derived using the asymptotic expansion technique.
 \end{abstract}

\end{frontmatter}

\setcounter{equation}{0}
\section{Introduction}
\noindent
In the present paper we show that, in the long-wave limit,  small-but-finite waves propagating in a one-dimensional medium  made of  nonlocally and nonlinearly elastic material  satisfy the Camassa-Holm (CH) equation \cite{camassa}   and the fractional CH equation (see equation (\ref{eq-s-f}))
 when a proper balance between dispersion and nonlinearity exists.

The CH equation
 \begin{equation}
      v_{\tau}+\kappa_{1} v_{\zeta}+3 vv_{\zeta}-v_{\zeta\zeta\tau}=\kappa_{2}(2 v_{\zeta}v_{\zeta\zeta}+vv_{\zeta\zeta\zeta}), \label{c-h}
 \end{equation}
was derived for the propagation of unidirectional small-amplitude shallow-water waves \cite{camassa,johnson1,constantin1, johnson2, lannes,ionescu} when the nonlinear effects are stronger than the dispersive effects. Due to the fact that, even for smooth initial data, the solution of the CH equation stays bounded as its slope becomes unbounded, it is often used as an appropriate model capturing the essential features of wave-breaking of shallow water waves  \cite{constantin2}. However, recalling that   (\ref{c-h})  is derived  under the long wavelength assumption, it follows that the CH equation is valid only when the solutions and their derivatives remain bounded \cite{lannes}. For a discussion on a different criterion for wave-breaking in long wave models we refer the reader to \cite{bjorkavag}. It is interesting to note that the CH equation as a model for wave-breaking of water  waves is an infinite-dimensional completely integrable Hamiltonian system \cite{constantin3, constantin4}.   Another interesting property of the CH equation is the existence of  the so-called peakon solitary wave solutions  when $\kappa_{1}=0$  \cite{camassa}.   At this point, it is worth pointing out that the derivation in the present study is also based on the long wavelength assumption and that  $\kappa_{1}$ is nonzero for the resulting equation. In addition to the studies about water waves, there are also studies that derive the CH equation as an appropriate model equation for nonlinear dispersive elastic waves. We refer the reader to \cite{dai} for the derivation of a CH-type equation governing the propagation of long waves in a compressible hyperelastic rod, and to \cite{chen} for the derivation of a two-dimensional CH-type equation governing the propagation of long waves in a compressible hyperelastic plate. However, these studies relied only on the "geometrical" dispersion resulting from the existence of the boundaries, that is, from the existence of a bounded elastic solid, like a rod or a plate. Another type of dispersion for elastic waves is the "physical" dispersion produced by the internal structure of the medium. Therefore, one interesting question is to investigate whether the CH equation can be derived as an asymptotic approximation for physically dispersive nonlinear elastic waves in the absence of the geometrical dispersion. In this study, we consider the one-dimensional wave propagation in an infinite, nonlinearly and  nonlocally elastic medium  whose constitutive behavior is described by a convolution integral. We then show that, for an exponential-type  kernel function,  the CH equation can model the propagation of  elastic waves even in the absence of the geometrical dispersion. Furthermore, by considering a fractional-type kernel function we are able to derive a fractional-type CH equation,  which indicates the possibility of obtaining more general evolution equations for suitable kernel functions. It is well-known that the KdV and BBM equations are valid at the same level of approximation while the CH equation is more accurate than the KdV and BBM equations. Therefore, when we neglect the highest order terms in the asymptotic expansion, the KdV and BBM equations and their fractional generalizations are also obtained as a by-product of the present derivation.  We underline that the asymptotic derivation of the CH equation needs a double asymptotic expansion in two small parameters characterizing nonlinear and  dispersive effects.  However, assuming simply that the two parameters are equal, the asymptotic derivations of the KdV and BBM equations can also be  based on a single asymptotic expansion in one small parameter resulting from the balance of nonlinear and dispersive effects.

The paper is organized as follows.  Section 2 presents the governing equations of one-dimensional nonlocal nonlinear elasticity theory and gives the equation of motion in dimensionless quantities for various forms of the kernel function. In Section 3, using a multiple scale asymptotic expansion, the CH equation is derived from the improved Boussinesq (IBq) equation  which is the equation of motion for the exponential kernel function.  Section 4 presents the derivation of a fractional CH equation from the equation of motion corresponding to a fractional-type kernel function.

 \setcounter{equation}{0}
 \section{A One-Dimensional Nonlinear Theory of Nonlocal Elasticity}
\noindent

We consider a one-dimensional, infinite, homogeneous, elastic medium with a nonlinear and nonlocal stress-strain relation (see \cite{eringen1,duruk1, duruk2} and the references cited therein for a more detailed discussion of the nonlocal model). In the absence of body forces the equation of motion is
\begin{equation}
    \rho_{0}u_{tt}=(S(u_{X}))_{X}, \label{motionu}
\end{equation}
where the scalar function $u(X,t)$ represents the displacement of a reference point $X$ at time $t$, $\rho_{0}$ is the mass density of the medium, $S=S(u_{X})$ is the stress and the subscripts denote partial derivatives. In contrast with classical elasticity, we take the constitutive equation for the stress $S$ as a general nonlinear and nonlocal function of the strain   $u_{X}$. That is, we assume that the stress at a reference point is a nonlinear function of the strain at all points  in the body. As in \cite{duruk1, duruk2}, the constitutive equation  of the present model has the form
\begin{equation}
    S(X,t)=\int_{\Bbb R} \alpha(|X-Y|)\sigma(Y,t)\mbox{d}Y, ~~~~~~~~~~~
    \sigma(X,t)=W^{\prime}(u_{X}(X,t)) \label{const}
\end{equation}
where  $\sigma$ is the   classical (local) stress, $W$ is the strain-energy density function, $Y$ denotes a generic point of the medium, $\alpha$ is a kernel function to be specified below, and the symbol $^\prime$ denotes differentiation. The kernel $\alpha$  acts as a weight function  that determines the relative contribution of the local stress $\sigma(Y,t)$ at a point $Y$ in a neighborhood of $X$ to the nonlocal stress $S(X,t)$. So, when the kernel becomes the Dirac delta function, the classical constitutive relation of a hyperelastic material is recovered.  Assuming the reference configuration is a stress-free undistorted state, we require that $W(0)=W^{\prime}(0)=0$.   We point out that if we take $W(u_{X})=(\lambda +2\mu)(u_{X})^{2}/2$ where $\lambda$ and $\mu$ are Lame constants, the above equations reduce to those of the linear theory of one-dimensional nonlocal elasticity (see \cite{eringen1}).

Without loss of generality, for convenience, the strain-energy density function may be considered to consist of a quadratic part $(u_{X})^{2}/2$ and a non-quadratic part $G(u_{X})$ with $G(0)=G^{\prime}(0)=0$:
\begin{displaymath}
    W(u_{X})=\gamma [\frac{1}{2}(u_{X})^{2}+G(u_{X})],
\end{displaymath}
where $\gamma$ is a constant with the dimension of stress. Differentiating both sides of (\ref{motionu}) with respect to $X$ and using (\ref{const}) we obtain the equation  of motion for the strain:
\begin{equation}
    \rho_{0}u_{Xtt}=\gamma\{ \int_{\Bbb R}
         \alpha(|X-Y|)[u_{X}+g(u_{X})]\mbox{d}Y\}_{XX}, \label{strain}
\end{equation}
where $g(s)=G^{\prime}(s)$ with $g(0)=0$. Now we define the dimensionless independent variables
\begin{displaymath}
    x={X\over l}, ~~~~~~\eta={t\over l}\sqrt{\gamma\over \rho_{0}}
\end{displaymath}
where $l$ is a characteristic length and from now on, and for simplicity, we use $u$ for $u_{X}$ and $t$ for $\eta$. Thus, (\ref{strain}) takes the  form
\begin{equation}
     u_{tt}=(\beta \ast (u+g(u)))_{xx}, \label{nonlocal}
\end{equation}
where the convolution operator $\ast$ is defined by
\begin{displaymath}
    \beta \ast v= \int_{\Bbb R} \beta(x-y)v(y)\mbox{d}y
\end{displaymath}
and $\beta(x)=l\alpha(|x|)$.   From a wave propagation point of view, the harmonic wave solutions to the linearized form of (\ref{nonlocal}) are dispersive  and  the sole source of dispersion in the present model is the internal structure of the medium but not the existence of the boundaries.
In general the kernel function $\beta$ is even, nonnegative and monotonically decreasing for $x>0$ (we refer the reader to \cite{eringen1} for the properties that an admissible kernel function must satisfy). A list of the most commonly used kernel functions is given in \cite{duruk2}. Here we consider two  kernel functions: the exponential kernel \cite{eringen2}  which is  the most widely used kernel function in the engineering applications of nonlocal elasticity \cite{eringen1,narendar}, and a fractional-type kernel function.  These two kernels are chosen because they are the simplest representatives of the kernels that are convenient for asymptotic expansions. Moreover, as discussed in Remark \ref{rem4.2}, starting from (\ref{nonlocal}) with a general kernel satisfying some mild assumptions will lead to the same results with those of the two representative kernels.

The exponential kernel is given by $\beta(x)={\frac{1}{2}}e^{-|x|}$.  The Fourier transform of $\beta$ is $\widehat{\beta}(\xi )=(1+\xi^{2})^{-1}$ where $\xi$ is the Fourier variable. Note that $\beta(x) $ is the Green's function for the operator  $~1-D_{x}^{2}~$ where $D_{x}$ represents the partial derivative with respect to $x$. Now, using the convolution theorem stating that the Fourier transform of the convolution of two functions is the product of their  Fourier transforms, we  take the Fourier transform of both sides of (\ref{nonlocal}). Then, substituting $\widehat{\beta}(\xi)$ of the exponential kernel into the resulting equation and taking the inverse Fourier transform, we obtain the equation of motion corresponding to the exponential kernel. Thus, for the exponential kernel, the equation of motion, (\ref{nonlocal}), reduces to  the IBq equation
\begin{equation}
    u_{tt}-u_{xx}-u_{xxtt}=(g(u))_{xx}.  \label{imbq}
\end{equation}
We next consider a fractional-type kernel function whose Fourier transform is $\widehat{\beta}(\xi )=(1+(\xi^{2})^{\nu})^{-1}$ where $\nu$ may not be an integer. Note that the previous case corresponds to $\nu=1$. To ensure the local well-posedness of the Cauchy initial-value problem defined for the resulting form of the equation of motion \cite{duruk2}, we impose the condition $\nu \geq 1$.  Noting that $\beta(x)$ is the Green's function for the operator   $~1+(-D_{x}^{2})^{\nu}$, this time the equation of motion, (\ref{nonlocal}), becomes an improved Boussinesq equation of fractional type
\begin{equation}
    u_{tt}-u_{xx}+(-D_{x}^{2})^{\nu}u_{tt}=(g(u))_{xx}~.  \label{frac}
\end{equation}
Here the operator $(-D_{x}^{2})^{\nu}$ is defined as $(-D_{x}^{2})^{\nu}q={\cal F}^{-1}(|\xi|^{2\nu}{\cal F}q)$ where  ${\cal F}$ and ${\cal F}^{-1}$ denote  the Fourier transform and its inverse, respectively. In Sections 3 and 4,  starting from (\ref{imbq}) and  (\ref{frac}), respectively, and restricting our attention to quadratic nonlinearities, we will investigate  the wave equations describing the unidirectional propagation of small-but-finite amplitude long waves. Knowing how the operator $(-D_{x}^{2})^{\nu}$ scales under the scaling  transformation $X=\delta x$ where $\delta$ is a positive constant will be important for the calculations in Section 4. To clarify the situation, let us consider the transformation $q(x)=Q(X)$. Then the relationship between the Fourier transforms of $q(x)$ and $Q(X)$ is
\begin{equation}
    \widehat{q}(\xi)=\int_{\Bbb R}e^{-ix\xi}q(x)dx={1\over \delta}\int_{\Bbb R}e^{-iX{\xi\over \delta}}Q(X)dX
    ={1\over \delta}\widehat{Q}({\xi\over \delta}).
\end{equation}
Thus, using the inverse Fourier transform we get
\begin{eqnarray}
    (-D_{x}^{2})^{\nu}q(x)&=&{1\over {2\pi\delta}}\int_{\Bbb R}|\xi|^{2\nu}e^{ix\xi}\widehat{Q}({\xi\over \delta})d\xi
            ={\delta^{2\nu}\over {2\pi}}\int_{\Bbb R} |K|^{2\nu}e^{iKX}\widehat{Q}(K)dK \nonumber \\
                &=&\delta^{2\nu}(-D_{X}^{2})^{\nu}Q(X).
\end{eqnarray}
We also note that the operator  $(-D_{x}^{2})^{\nu}$ is a translation invariant operator, so that $(-D_{x}^{2})^{\nu}(q(x+\gamma))=((-D_{x}^{2})^{\nu}q)(x+\gamma)$ where $\gamma$ is a constant.

 \setcounter{equation}{0}
 \section{Derivation of The Camassa-Holm Equation in the Long Wave Limit}
\noindent
In this section we provide a formal derivation of the CH equation from (\ref{imbq}) with $g(u)=u^{2}$ in the long wave limit. We restrict our attention to the quadratic nonlinearity since the KdV, BBM and CH equations all contain only quadratic nonlinearities.   For a data on a compact support, (\ref{imbq}) has both right-going and left-going wave solutions that are moving  apart.  Assuming the two waves no longer overlap for a sufficiently large time, in the rest of this section  we consider  right- going, small-but-finite amplitude, long wave solutions of (\ref{imbq}). We first introduce the scaling transformation
 \begin{equation}
    u(x,t)= \epsilon U(\delta (x-t),\delta t)=\epsilon U(Y,S) \label{approx}
 \end{equation}
with $Y=\delta (x-t)$ and $S=\delta t$  to make the asymptotic behavior of (\ref{imbq}) more transparent. Here, small parameters  $\epsilon>0$ and $\delta>0$ measure nonlinear and dispersive effects, respectively; more precisely, $\epsilon$ denotes a typical (small) amplitude of waves whereas $\delta$ denotes a typical (small) wavenumber, which is equivalent to supposing that the waves are long.  Following the approaches in \cite{johnson1,constantin1,johnson2,lannes} we seek a solution in the form of a double asymptotic expansion in two small parameters $\epsilon$ and $\delta$. The double limit process  allows us to  control dispersive and nonlinear effects  in capturing the CH equation. Note that the asymptotic derivations of the KdV and BBM equations can be  based on a single asymptotic expansion in one small parameter by taking, for instance,  $\delta^{2}=\epsilon$.  Inserting the scale transformation (\ref{approx}) into  (\ref{imbq})  and multiplying  the resulting equations by  $\epsilon^{-1} \delta^{-2}$ lead to the equation
 \begin{equation}
     U_{SS}- 2U_{YS}  -\delta^2(U_{YYYY}+U_{YYSS}-2U_{YYYS}) -\epsilon  (U^2)_{YY}=0, \label{per-bous}
 \end{equation}
for $U(Y,S)$. We now seek an asymptotic solution of (\ref{per-bous}) in the form
 \begin{equation}
    U(Y,S;\epsilon, \delta)= U_{0}(Y,S)+\epsilon U_{1}(Y,S)+\delta^2 U_{2}(Y,S)+\epsilon \delta^2 U_{3}(Y,S) +{\cal O}(\epsilon^{2},\delta^{4}) \label{sol}
 \end{equation}
 as $\epsilon \rightarrow 0$, $\delta \rightarrow 0$. Note that odd powers of $\delta$ will never appear due to the existence of even order spatial derivatives only in the IBq equation. We assume that the unknowns $U_{n}$ $(n=0,1,2, ...)$ and their derivatives decay to zero as  $|Y| \rightarrow \infty$.  Substituting the approximate solution (\ref{sol}) into (\ref{per-bous}) we obtain a hierarchy
 of equations  in powers of  $\epsilon$ and $\delta^{2}$. The leading-order equation is given by
 \begin{equation}
    (D_S-2D_Y) U_{0S} =0. \label{eq-a}
 \end{equation}
Recalling that we have already restricted our attention to the right-going solutions we take $U_{0S}=0$ which implies  $U_{0}=U_{0}(Y)$. At  ${\cal O}(\epsilon)$, we have
 \begin{equation}
    (D_S-2D_Y) U_{1S}  -((U_{0})^2)_{YY}=0. \label{eq-b}
 \end{equation}
 Differentiating this equation with respect to $S$ we get an equation for $U_{1SS}$ which is the same as (\ref{eq-a}). With a similar argument, we take $U_{1SS}=0$. Then the solution of (\ref{eq-b}) is obtained as
 \begin{equation}
    U_{1S}=-\frac{1}{2}((U_{0})^2)_Y.  \label{eq-c}
 \end{equation}
At ${\cal O}(\delta^2)$, we have
 \begin{equation}
    (D_S-2D_Y) U_{2S} -U_{0YYYY}=0, \label{eq-d}
 \end{equation}
 where  we have used the fact $U_{0S}=0$. Differentiating this equation with respect to $S$ and then following the same line of reasoning,  we take $U_{2SS}=0$. Then the solution of (\ref{eq-d}) is
 \begin{equation}
    U_{2S}=-\frac{1}{2}U_{0YYY}. \label{eq-e}
 \end{equation}
 Finally, at ${\cal O}(\epsilon \delta^2)$, the equation to  be satisfied is of the form
 \begin{equation}
    (D_S-2D_Y) U_{3S}  -U_{1YYYY}+2 U_{1YYYS} -2(U_{0} U_{2})_{YY}=0, \label{eq-f}
  \end{equation}
 where the fact $U_{1SS}=0$ has been used. Differentiating this equation twice with respect to $S$ we get an equation for $U_{3SSS}$ which is the same as (\ref{eq-a}). To follow the right-going wave, again we take $U_{3SSS}=0$. Differentiating (\ref{eq-f}) with respect to $S$ and using   (\ref{eq-c}), (\ref{eq-e}) and  $U_{3SSS}=0$, we get
 \begin{equation}
    U_{3SS}=\frac{1}{4} ((U_{0})^2)_{YYYY}+\frac{1}{2} (U_{0} U_{0YYY})_Y. \label{eq-g}
 \end{equation}
 Substituting this result into  (\ref{eq-f}) gives $U_{3S}$ in the form
 \begin{equation}
    U_{3S}=-(U_{0} U_{2})_Y-\frac{1}{2} U_{1YYY}-\frac{3}{4} (U_{0}U_{0Y})_{YY}+\frac{1}{4} U_{0} U_{0YYY}. \label{eq-k}
 \end{equation}
Combining the above results with (\ref{sol}) we obtain
 \begin{eqnarray}
    U_S &=&\epsilon  U_{1S} +\delta^2  U_{2S}+ \epsilon\delta^2  U_{3S}+{\cal O}(\epsilon^{2},\delta^{4}) \nonumber\\
        &=& -\frac{\epsilon}{2}  \left[ (U_{0})^2 + 2\delta^2 U_{0} U_{2}\right]_Y
            -\frac{\delta^2}{2} \left(U_{0} + \epsilon U_{1}\right)_{YYY} \nonumber \\
        &~&    -\frac{\epsilon \delta^2}{4} (9 U_{0Y} U_{0YY} + 2 U_{0} U_{0YYY})+{\cal O}(\epsilon^{2},\delta^{4}). \label{eq-h}
 \end{eqnarray}
At this level of approximation, adding $U_{0S}=0$, (\ref{eq-c}) and  (\ref{eq-e}) to (\ref{eq-h}),  we form the equation for $U(Y,S;\epsilon, \delta)$ as follows
 \begin{equation}
    U_S+\epsilon UU_Y +\frac{\delta^2}{2} U_{YYY}
        +\frac{\epsilon \delta^2}{4} (9 U_Y U_{YY} + 2 U U_{YYY})=0. \label{eq-i}
 \end{equation}
We note that this equation is not in the standard CH form, for instance, the term $U_{YYS}$ of the CH equation is absent. But this can be remedied if we rewrite (\ref{eq-i}) in a moving frame defined by
\begin{equation}
    X=aY+bS, ~~~~T=cS \label{eq-j}
\end{equation}
where $a$, $b$ and $c$ are positive constants. The free parameters $a$, $b$ and $c$ will be chosen later so that (\ref{eq-i}) becomes the CH equation in the new frame defined by $X$ and $T$. Rewriting (\ref{eq-i}) in the new coordinate system with $U(Y,S)=V(aY+bS, cS)=V(X,T)$ we obtain
\begin{equation}
 cV_T+b V_X+\epsilon a VV_X +\frac{\delta^2 a^{3}}{2} V_{XXX} +\frac{\epsilon \delta^2  a^{3}}{4} (9 V_X V_{XX} + 2 V V_{XXX})=0. \label{eq-k}
\end{equation}
This equation still lacks a term corresponding to $v_{\tau\zeta\zeta}$ in (\ref{c-h}). To remove the term $V_{XXX}$ in favor of $V_{TXX}$, as in \cite{johnson1,constantin1,johnson2}, we use
 \begin{equation}
  V_{XXX}=-{c\over b}V_{TXX}-{a\over b}\epsilon (VV_{X})_{XX}+{\cal O}(\delta^{2},\epsilon\delta^{2}), \label{eq-l}
 \end{equation}
obtained from  (\ref{eq-k}), which is valid at this level of approximation. Thus, (\ref{eq-k}) takes the following form
 \begin{equation}
  V_T+{b\over c} V_X+{{\epsilon a}\over c} VV_X -\frac{\delta^2 a^{3}}{2b} V_{TXX}+\frac{\epsilon \delta^2 a^{3}}{4c}[3(3-\frac{2a}{b}) V_X V_{XX} + 2(1-\frac{a}{b}) V V_{XXX}]=0.  \label{eq-m}
 \end{equation}
 To remove the parameters $\epsilon$ and $\delta$ from this equation we introduce the scaling transformation
 \begin{equation}
    v=\epsilon V, ~~~~X=\delta \zeta, ~~~~T=\delta \tau. \label{eq-n}
\end{equation}
Then (\ref{eq-m}) becomes
\begin{equation}
  v_\tau+{b\over c} v_\zeta+{a\over c} vv_\zeta -\frac{a^{3}}{2b} v_{\tau\zeta\zeta}+\frac{a^{3}}{4c}[3(3-\frac{2a}{b}) v_\zeta v_{\zeta\zeta} + 2(1-\frac{a}{b}) v v_{\zeta\zeta\zeta}]=0.  \label{eq-o}
 \end{equation}
A special form of the CH equation given by (\ref{c-h}), which corresponds to the special values of $\kappa_{1}$ and $\kappa_{2}$,  can now be recovered by making suitable choices for the parameters $a$, $b$ and $c$. For a standard CH equation,  the ratio of the coefficients of the terms $v_\zeta v_{\zeta\zeta}$ and $v v_{\zeta\zeta\zeta}$ is important and it must be 2:1, respectively \cite{johnson1,constantin1,johnson2} (see \cite{lannes} for a detailed discussion on this issue).  In order to recover (\ref{c-h}) we also impose the conditions that the coefficients of the terms $vv_{\zeta}$ and $v_{\tau\zeta\zeta}$ are 3 and -1, respectively. To ensure that all three conditions are satisfied we require
\begin{equation}
a={2\over \sqrt{5}},~~~~b={4\over {5\sqrt{5}}},~~~~c={2\over {3\sqrt{5}}}. \label{param-a}
 \end{equation}
With this choice of the parameters,  (\ref{eq-m}) and (\ref{eq-o}) reduce to
\begin{equation}
        V_{T}+\frac{6}{5}V_{X}+3\epsilon VV_{X}-\delta^{2}V_{TXX}-\frac{9\epsilon \delta^{2}}{5}(2V_{X}V_{XX}+VV_{XXX})=0, \label{eq-r}
\end{equation}
and
\begin{equation}
      v_{\tau}+{6\over 5} v_{\zeta}+3 vv_{\zeta}-v_{\zeta\zeta\tau}={9\over 5}(2 v_{\zeta}v_{\zeta\zeta}+vv_{\zeta\zeta\zeta}), \label{eq-s}
 \end{equation}
respectively. We note that (\ref{eq-s}) is the standard CH equation (\ref{c-h}) with $\kappa_{1}={6/5}$ and $\kappa_{2}={9/5}$.

From (\ref{approx}), (\ref{eq-j}), (\ref{eq-n}) and (\ref{param-a}) it follows that the coordinate transformation between $(\zeta, \tau)$ and $(x,t)$ is
\begin{equation}
    \zeta={2\over \sqrt{5}}(x-{3\over 5}t), ~~~~ \tau={2\over {3\sqrt{5}}}t.  \label{trans}
\end{equation}
By applying this coordinate transformation, we can write (\ref{eq-s}) in the original reference frame as follows
\begin{equation}
      v_{t}+ v_{x}+ vv_{x}-{3\over 4}v_{xxx}-{5\over 4}v_{xxt}={3\over 4}(2 v_{x}v_{xx}+vv_{xxx}) \label{ch-or}
 \end{equation}
with $v=v(x, t)$.

\begin{remark}\label{rem3.1}
We point out that if we neglect the terms of order $\epsilon \delta^2$ and higher in (\ref{sol}) we obtain the Benjamin-Bona-Mahony (BBM) equation \cite{bbm}, instead of (\ref{eq-s}). In other words, if we seek an asymptotic solution of (\ref{per-bous}) in the form
 \begin{equation}
    U(Y,S;\epsilon, \delta)= U_{0}(Y,S)+\epsilon U_{1}(Y,S)+\delta^2 U_{2}(Y,S) +{\cal O}(\epsilon^{2},\epsilon \delta^2, \delta^{4}) \label{sol-a}
 \end{equation}
and if we follow similar steps in this section, we get the Korteweg-de Vries (KdV) \cite{korteweg} equation
 \begin{equation}
    U_S+\epsilon UU_Y +\frac{\delta^2}{2} U_{YYY}=0 \label{eq-i-a}
 \end{equation}
instead of  (\ref{eq-i}) and then the BBM equation
\begin{equation}
      v_{\tau}+\kappa_{1} v_{\zeta}+3 vv_{\zeta}-v_{\zeta\zeta\tau}=0, \label{eq-s-a}
 \end{equation}
with $\kappa_{1}=3a^{2}/2$ (where $a$ is an arbitrary positive constant), instead of (\ref{eq-s}). In the original reference frame $(x,t)$, (\ref{eq-i-a}) and (\ref{eq-s-a})  are written  in the form
 \begin{equation}
      v_{t}+ v_{x}+ vv_{x}+{1\over 2}v_{xxx}=0 \label{kdv-or}
 \end{equation}
and
 \begin{equation}
      v_{t}+ v_{x}+ vv_{x}-{3\over 4}v_{xxx}-{5\over 4}v_{xxt}=0 \label{bbm}
 \end{equation}
with $v=v(x, t)$, respectively.
\end{remark}

  \setcounter{equation}{0}
 \section{Derivation of a Fractional Camassa-Holm Equation}

In this section we will handle (\ref{frac}) with $g(u)=u^{2}$ and explore how to extend the asymptotic expansion of the previous section to (\ref{frac}). As we did in the previous section, to make the asymptotic behavior of (\ref{frac}) more transparent, we use the scaling transformation (\ref{approx}) in (\ref{frac}) and we get
 \begin{equation}
     U_{SS}- 2U_{YS}  +\delta^{2\nu}(-D_{Y}^{2})^{\nu}(U_{SS}+U_{YY}-2U_{SY}) -\epsilon  (U^2)_{YY}=0. \label{per-bous-f}
 \end{equation}
We then seek an asymptotic solution of (\ref{per-bous-f}) in the form
 \begin{equation}
    U(Y,S;\epsilon, \delta)= U_{0}(Y,S)+\epsilon U_{1}(Y,S)+\delta^{2\nu} U_{2}(Y,S)+\epsilon \delta^{2\nu} U_{3}(Y,S) +{\cal O}(\epsilon^{2},\delta^{4\nu}) \label{sol-f}
 \end{equation}
 with the assumptions we made for (\ref{sol}). Substitution of  (\ref{sol-f}) into (\ref{per-bous-f}) leads to a hierarchy  of equations  in powers of  $\epsilon$ and $\delta^{2\nu}$. As the process is quite similar to that of the previous section, we  outline  the basic steps involved in the asymptotic derivation. Again, a solution of the leading order equation is $U_{0}=U_{0}(Y)$. Similarly, at ${\cal O}(\epsilon)$, the asymptotic equation admits the solution given by (\ref{eq-c}). At ${\cal O}(\delta^{2\nu})$, the asymptotic equation admits the solution
 \begin{equation}
    U_{2S}=\frac{1}{2}(-D_{Y}^{2})^{\nu}U_{0Y}. \label{eq-e-f}
 \end{equation}
 At ${\cal O}(\epsilon \delta^{2\nu})$,  the equation to  be satisfied is of the form
 \begin{equation}
    (D_S-2D_Y) U_{3S}  +(-D_{Y}^{2})^{\nu}(U_{1YY}-2 U_{1SY}) -2(U_{0} U_{2})_{YY}=0, \label{eq-f-f}
  \end{equation}
 instead of (\ref{eq-f}). Following the same line of argument we obtain the solution of (\ref{eq-f-f}) as
  \begin{equation}
    U_{3S}=-(U_{0} U_{2})_Y+\frac{1}{2}(-D_{Y}^{2})^{\nu} U_{1Y}+\frac{3}{4}(-D_{Y}^{2})^{\nu}(U_{0}U_{0Y})-\frac{1}{4} U_{0}(-D_{Y}^{2})^{\nu} U_{0Y}. \label{eq-k-f}
 \end{equation}
 Substituting the above results  for $U_{1S}$, $U_{2S}$ and $U_{3S}$  into
 \begin{equation}
    U_S =\epsilon  U_{1S} +\delta^{2\nu}  U_{2S}+ \epsilon\delta^{2\nu}  U_{3S}+{\cal O}(\epsilon^{2},\delta^{4\nu})
         \label{eq-h-f}
 \end{equation}
 we get
  \begin{equation}
    U_S+\epsilon UU_Y -\frac{\delta^{2\nu}}{2}(-D_{Y}^{2})^{\nu} U_{Y}
        -\frac{\epsilon \delta^{2\nu}}{4} [3 (-D_{Y}^{2})^{\nu} ( U U_{Y})- U  (-D_{Y}^{2})^{\nu} U_{Y}]=0. \label{eq-i-f}
 \end{equation}
 for $U(Y,S;\epsilon, \delta)$, which is valid at this level of approximation. Using the coordinate transformation (\ref{eq-j}), we rewrite (\ref{eq-i-f}) in the new coordinate system with $U(Y,S)=V(aY+bS, cS)=V(X,T)$ and then remove the term $(-D_{X}^{2})^{\nu} V_{X}$ in favor of $(-D_{X}^{2})^{\nu} V_{T}$:
\begin{eqnarray}
 && V_T+{b\over c} V_X+{\epsilon a\over c} VV_X +\frac{\delta^{2\nu} a^{2\nu+1}}{2b}(-D_{X}^{2})^{\nu} V_{T}  \nonumber \\
 &&  ~~~~~~~~~~~~~    +\frac{\epsilon \delta^{2\nu} a^{2\nu+1}}{4c}[(\frac{2a}{b}-3) (-D_{X}^{2})^{\nu}(V V_{X}) +  V  (-D_{X}^{2})^{\nu} V_{X}]=0.
                     \label{eq-m-f}
 \end{eqnarray}
 Using the scale transformation  (\ref{eq-n}) and following the same line of argument that led to  (\ref{eq-r}) and (\ref{eq-s}), we obtain
 \begin{equation}
        V_{T}+\frac{6}{5}V_{X}+\frac{3\epsilon}{2}(V^{2})_{X}+\delta^{2\nu}(-D_{X}^{2})^{\nu}V_{T}+\frac{3\epsilon \delta^{2\nu}}{5}[ 2(-D_{X}^{2})^{\nu}(VV_{X})+V (-D_{X}^{2})^{\nu}  V_{X}]=0, \label{eq-r-f}
\end{equation}
instead of  (\ref{eq-r})  and the fractional CH equation
\begin{equation}
      v_{\tau}+{6\over 5} v_{\zeta}+3 vv_{\zeta}+(-D_{\zeta}^{2})^{\nu}v_{\tau}=-{3\over 5}[2 (-D_{\zeta}^{2})^{\nu}  (vv_{\zeta})+v  (-D_{\zeta}^{2})^{\nu}v_{\zeta}], \label{eq-s-f}
 \end{equation}
 instead of (\ref{eq-s}). In the present case, the parameters $a$, $b$ and $c$ are obtained as
 \begin{equation}
a=({4\over 5})^{1/2\nu},~~~~b={2\over 5}a,~~~~c={a\over 3}. \label{param-b}
 \end{equation}
To the best of our knowledge, the fractional CH equation, (\ref{eq-s-f}), has never appeared before in the literature.

From (\ref{approx}), (\ref{eq-j}), (\ref{eq-n}) and (\ref{param-b})  we get the coordinate transformation between $(\zeta, \tau)$ and $(x,t)$ as
\begin{equation}
    \zeta=a(x-{3\over 5}t), ~~~ \tau={a\over 3}t.  \label{trans-a}
\end{equation}
Using this coordinate transformation,   (\ref{eq-s-f}) can be rewritten as
 \begin{equation}
        v_{t}+ v_{x}+ vv_{x}+{3\over 4}(-D_{x}^{2})^{\nu}v_{x}+{5\over 4}(-D_{x}^{2})^{\nu}v_{t}=-{1\over 4}[2  (-D_{x}^{2})^{\nu}   (vv_{x})+v (-D_{x}^{2})^{\nu}  v_{x}]
       \label{eq-s-f-or}
 \end{equation}
for $v=v(x, t)$  in the original reference frame.

 \begin{remark}\label{rem4.1}
As in the previous section,  if we neglect the terms of order $\epsilon \delta^{2\nu}$ and higher in (\ref{sol-f}) we obtain a fractional BBM equation, instead of (\ref{eq-s-f}). That is, if we seek an asymptotic solution of (\ref{per-bous-f}) in the form
 \begin{equation}
    U(Y,S;\epsilon, \delta)= U_{0}(Y,S)+\epsilon U_{1}(Y,S)+\delta^{2\nu} U_{2}(Y,S) +{\cal O}(\epsilon^{2},\epsilon \delta^{2\nu}, \delta^{4\nu}) \label{sol-a-f}
 \end{equation}
and if we follow similar steps in this section, we get the fractional KdV equation \cite{bona}
 \begin{equation}
    U_S+\epsilon UU_Y -\frac{\delta^{2\nu}}{2}(-D_{Y}^{2})^{\nu} U_{Y}=0
         \label{eq-i-ff}
 \end{equation}
instead of  (\ref{eq-i-f}) and then the fractional BBM equation \cite{kapitula}
\begin{equation}
      v_{\tau}+\kappa_{1} v_{\zeta}+3 vv_{\zeta}+(-D_{\zeta}^{2})^{\nu}v_{\tau}=0,  \label{eq-s-ff}
 \end{equation}
 with $\kappa_{1}=3a^{2\nu}/2$ (where $a$ is an arbitrary positive constant), instead of (\ref{eq-s-f}). In the original reference frame $(x,t)$, (\ref{eq-i-ff}) and (\ref{eq-s-ff})  take the following forms
 \begin{equation}
     v_{t}+ v_{x}+ vv_{x}-{1\over 2}(-D_{x}^{2})^{\nu} v_{x}=0
         \label{fkdv-or}
 \end{equation}
and
 \begin{equation}
       v_{t}+ v_{x}+ vv_{x}+{3\over 4}(-D_{x}^{2})^{\nu}v_{x}+{5\over 4}(-D_{x}^{2})^{\nu}v_{t}=0
      \label{fbbm-or}
 \end{equation}
with $v=v(x, t)$, respectively.
\end{remark}

 \begin{remark}\label{rem4.2}
In Sections 3 and 4, we have derived the CH, KdV and BBM equations and their fractional counterparts by starting with (\ref{nonlocal}) and considering the kernels defined by $\widehat{\beta}(\xi )=\left(1+\xi^{2}\right)^{-1}$ and $\widehat{\beta}(\xi )=\left(1+(\xi^{2})^{\nu}\right)^{-1}$.  The natural question is what would happen if we had started with an arbitrary kernel function $\beta(x)$. To provide the answer to this question, let us suppose that the Fourier transform $\widehat{\beta}(\xi )$ is regular enough to have an expansion of the form, up to constants,
\begin{displaymath}
    1/\widehat{\beta}(\xi )=1+(\xi^{2})^{\nu}+{\cal O}\left((\xi^{2})^{\rho}  \right)
\end{displaymath}
with $1\leq \nu <\rho$ for $\xi$ near $0$.  When we consider an asymptotic expansion (like  (\ref{sol-f})) for a solution of (\ref{nonlocal}), we observe that the term ${\cal O}\left((\xi^{2})^{\rho}  \right)$ in the above equation yields to ${\cal O}\left(\delta^{2\rho}\right)$ terms. Thus, only the leading-order terms $1+(\xi^{2})^{\nu}$, corresponding to the operator $1+\left(-D_{x}^{2}\right)^{\nu}$, will be effective at the level of approximation considered in this study. In other words, up to constants, the asymptotic derivation will  give rise to exactly the same set of evolution equations, that is,  (\ref{eq-s-f}), (\ref{eq-i-ff}) and (\ref{eq-s-ff}).  In that sense our results in Sections 3 and 4 in fact hold for  (\ref{nonlocal}) with a more general class of kernels.
\end{remark}

\noindent

 \end{document}